\documentclass[reprint,amsmath,amssymb,aps,prb,floatfix,a4paper, superscriptaddress]{revtex4-1}

\usepackage{graphicx}
\usepackage{dcolumn}
\usepackage{bm}
\usepackage{color}
\usepackage[ugly]{nicefrac}
\usepackage[latin1]{inputenc}
\usepackage{subfig}
\usepackage{SIunits}
\usepackage{epstopdf}

\newcommand{\etal}{$\textit{et\ al.\ }$}

\begin{document}
\title{Revealing the spin and symmetry properties of the buried $\mathrm{Co_{2}MnSi}/MgO$ interface by low energy spin-resolved photoemission}
\author{Roman Fetzer}
\email{rfetzer@rhrk.uni-kl.de}
\author{Marcel L\"osch}
\affiliation{Department of Physics and Research Center OPTIMAS, University of Kaiserslautern, Erwin-Schr\"odingerstr.\,46, 67663 Kaiserslautern, Germany}
\author{Yusuke Ohdaira}
\author{Hiroshi Naganuma}
\author{Mikihiko Oogane}
\author{Yasuo Ando}
\affiliation{Department of Applied Physics, Graduate School of Engineering, Tohoku University, aoba-yama 6-6-05, Sendai 980-8579, Japan}
\author{Tomoyuki Taira}
\author{Tetsuya Uemura}
\author{Masafumi Yamamoto}
\affiliation{Division of Electronics for Informatics, Hokkaido University, Kita 14 Nishi 9, Sapporo 060-0814, Japan}
\author{Martin Aeschlimann}
\author{Mirko Cinchetti}
\affiliation{Department of Physics and Research Center OPTIMAS, University of Kaiserslautern, Erwin-Schr\"odingerstr.\,46, 67663 Kaiserslautern, Germany}

\date{\today}
\begin{abstract}

We present a novel approach to study the spin and symmetry electronic properties of buried interfaces using low-energy  spin-resolved photoemission spectroscopy. We show that this method is sensitive to interfaces buried below more than 20~ML ($\simeq$~4~nm) MgO, providing a powerful tool for the non-destructive characterization of spintronics interfaces. As a demonstration, we apply this technique to characterize the $Co_{2}MnSi/MgO$ interface, a fundamental building block of state-of-the-art magnetic tunnel junctions based on Heusler compounds. We find that a surface state with $\Delta_1$ symmetry and minority spin character dominating the electronic structure of the bare  $Co_{2}MnSi(100)$ surface  is quenched at the $Co_{2}MnSi(100)/MgO$ interface. As a result, the interface spin-dependent electronic structure resembles the theoretically expected $Co_{2}MnSi$ bulk band structure, with majority spin electronic states of both $\Delta_1$ and $\Delta_5$ symmetry. Furthermore we find an additional thermally-induced contribution in the minority channel, mirroring the $\Delta_1$/$\Delta_5$ asymmetry of the majority channel. 

\end{abstract}
\maketitle


Magnetic tunnel junctions (MTJs) are a typical example of exploiting spintronics concepts for industrial applications \cite{Moodera95, Chappert07}. A typical MTJ consists of two ferromagnetic (FM) electrodes spaced by an insulating material. If a finite voltage is applied to the structure, the resulting tunneling current  will depend on the relative magnetization direction of both electrodes, making such devices highly suitable as magnetic sensors e.g. in hard disk read heads. The occuring resistance change is called tunnel magnetoresistance (TMR) ratio and serves as the figure of merit for MTJs.\newline 
A promising approach to further improve the TMR ratio and hence the sensitivity and effectiveness of MTJs is the combination of half-metallic (HM) magnetic electrodes and crystalline barriers.
Utilizing half-metals as electrodes in MTJs should lead to a complete absence of a tunneling current for antiparallel magnetization alignment due to the lack of either initial or final states for one spin channel, resulting in a nearly infinite TMR ratio. Most successfully cobalt-based full Heusler compounds  like $Co_{2}MnSi$ (CMS) with a Curie Temperature of $\approx$~1000°C were applied as ferromagnetic half-metals in tunnel junctions together with amorphous $AlO_x$ barriers as well as advanced epitaxial MgO barriers~\cite{Sakuraba06a, Ishikawa08}. For the latter ones the tunnel probability depends on the electron eigenstate symmetry, i.e. it can lead to further increased TMR ratios for ferromagnetic materials with suitable band structure like Fe or Co~\cite{Butler01, Parkin04}. This should also be the case for CMS~\cite{Miura07}.\newline
Since the tunneling process involves only the outermost atomic layers of the ferromagnetic leads, the MTJ performance depends ultimately on the properties of the very ferromagnetic metal(FM)/insulator interface, more specifically on its spin polarization and on the symmetry of the electronic wave functions. As TMR devices have a typical multilayer structure, the relevant FM/insulator interfaces are not easy accessible. One possibility to study such interfaces is to use standard spin-resolved photoemission spectroscopy. However, the information depth of this method is limited by the extremely short electron mean free path, which is lower than 1~nm in MgO for common ultraviolet photoemission spectroscopy (UPS) and soft X-ray excitation sources (photon energies between 20 and 350~eV)~\cite{Nist00}. As a result only samples with ultrathin insulator layers - typically thinner than 2~ML in case of MgO - can be investigated on top of the ferromagnetic material~\cite{Matthes04, Plucinski07, Bonell12}. In such cases the presence of pinholes can never be completely ruled out, and as a consequence the studied interface is not identical to the interfaces buried in real devices. Going to even higher excitation energies photoemission spectroscopy becomes bulk-sensitive, hence no information about the interface itself can be obtained~\cite{Fecher08}. This disadvantage can be overcome partly by applying more sophisticated techniques like standing wave core and valence photoemission~\cite{Yang11}, which is a powerful tool to investigate most of the interface properties. Unfortunately, up to now it lacks the ability of spin detection. \newline 
Here we show that interfaces relevant for advanced spintronics devices can be directly studied by means of low-energy spin-resolved photoemission spectroscopy (SR-PES), taking advantage of the extremely large mean free path of electrons excited close to the conduction band edge of an insulating layer. We apply this novel method to characterize the CMS/MgO interface buried below epitaxial MgO films possessing the same thickness used in state-of-the-art MTJs ($\approx$~10~ML). At the CMS/MgO interface we observe the quenching of  a surface state with $\Delta_1$ symmetry and minority spin character, which dominates the electronic structure of the bare CMS surface in direct vicinity of the  Fermi energy. As a result, the interface spin-dependent electronic structure resembles the theoretically expected $Co_{2}MnSi$ bulk band structure showing predominant majority electron character from both $\Delta_1$ and $\Delta_5$ states, i.e. no additional interface states occur. Strikingly, the spin-polarization of the CMS/MgO interface is detectable even for much thicker MgO overlayers (up to more than 20~ML), demonstrating the power of low-energy SR-PES for the characterization of the spin and symmetry electronic properties of buried MgO interfaces. \newline  
Samples with the stacking structure MgO(001)sub/MgO(10\,nm)/CMS(50\,nm)/MgO(10\,ML $\cong$ 2\,nm) were fabricated~\cite{Tsunegi09a} to investigate the electron symmetry and spin polarization of the CMS/MgO interface as well as the free CMS surface. Additionally Mn-rich CMS(30\,nm)/MgO(20\,ML $\cong$ 4\,nm) samples served to prove the truly interface-sensitive property of low-energy SR-PES~\cite{Ishikawa09b}. The sample surface normal and hence the probing direction in all cases is the $\Gamma$-X crystalline direction with respect to CMS. Crystalline structure and chemical composition of the CMS/MgO bilayers as well as the CMS surface were controlled by low energy electron diffraction and Auger electron spectroscopy. The latter method is further used to ensure the complete removal of the MgO top layer by gentle 500~eV $Ar^{+}$ ion sputtering in order to investigate the bare CMS surface of the CMS/MgO(10\,ML) samples. Also a very short sputtering cycle with a removal of max. 1~ML of MgO was applied prior to photoemission experiments at the interface to remove residual carbon.  As expected, preferential sputtering at MgO did not take place~\cite{Henrich85}. In all cases the samples were subsequently annealed to 450°C for at least 30~minutes. Since the CMS/MgO(20\,ML) sample showed only marginal carbon adsorption at the surface due to less surface defects~\cite{Fetzer12a}, it was not $Ar^{+}$ ion sputtered. \newline
For the low-energy SR-PES experiments we used the fourth harmonic of a femtosecond Ti:Sa oscillator with $h\nu=5.9~eV$  as the light source. A $\frac{\lambda}{2}$ plate allows to switch between p- and s-polarization of the laser beam, which illuminates the sample  under a degree of 45°.  Regarding optical dipole selection rules and the sample surface orientation, the s-polarization setting will only excite electrons with $\Delta_5$ symmetry, while for p-polarization both $\Delta_1$ and $\Delta_5$ states are probed~\cite{Hermanson77, Bonell12}. By comparing the spectra obtained with the two light polarizations \cite{Wuestenberg12}  we can probe electronic states with $\Delta_1$ symmetry mainly responsible for the tunneling current in MgO-based MTJs~\cite{Butler01, Miura07}.  For comparison with standard photoemission methods, spin-resolved UPS (SR-UPS) spectra were recorded using a  commercial He VUV gas discharge lamp providing unpolarized light with $h\nu=21.2~eV$ (He I line). An Omicron CSA-SPLEED detector allows for energy- and spin-resolution of the photoemitted electrons. The energy resolution is set to 210~meV for laser excitation and 420~meV if the He lamp is used in order to account for the lower photoemission yield. All measurements were conducted at room temperature (RT) in a UHV chamber with a base pressure lower than $10^{-10}$~mbar. \newline
\begin{figure}
\centering
\includegraphics[width=\linewidth]{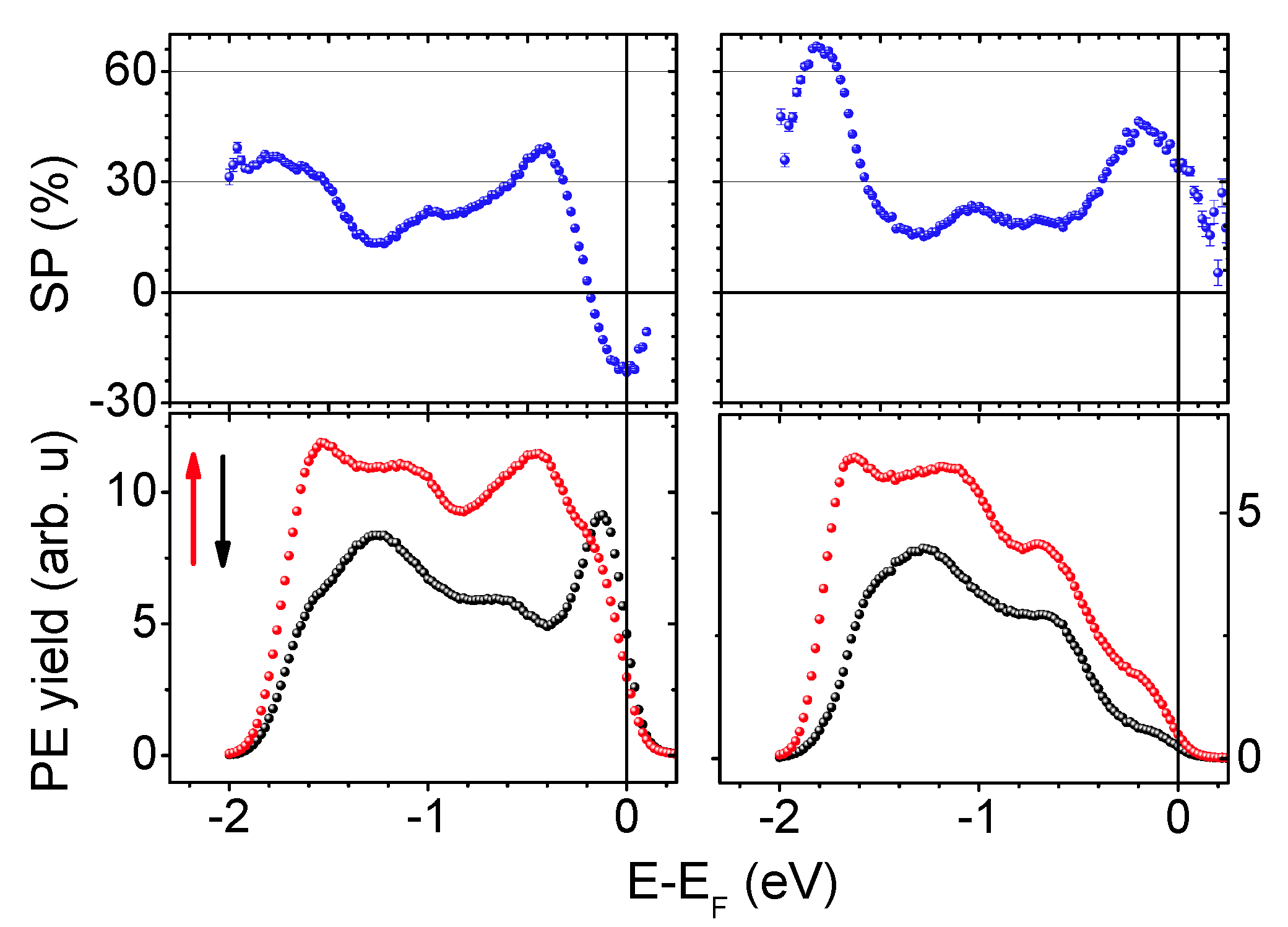}
\caption{Majority (red) and minority (black) electron photoemission spectra of the bare CMS(100) surface (lower panel) and deduced spin polarization (SP, upper panel), measured by low-energy SR-PES with s-polarized (left) and p-polarized laser light (right).}
\label{fig:surfaceSP}
\end{figure} 
For clarity we start discussing the low-energy SR-PES results obtained from the bare CMS surface. Figure \ref{fig:surfaceSP} shows the spin-resolved photoemission spectra of the free CMS(100) surface (lower panel) and the deduced spin polarization (upper panel), measured by low-energy SR-PES with p-polarized (left) and s-polarized laser light (right). For s-polarization the bulk CMS electron density of states (DOS) (c.f. Ref.~\cite{Balke06}) is reproduced quite well, with a low photoemission yield directly at the Fermi energy ($E_F$) originating from sp~states and distinct peaks originating from d~states dominating the spectra for binding energies below $-0.5$~eV. The spin polarization (SP) shows a maximum very near to $E_F$ at $E_B= -0.2~eV$ with a value of +45\%, possibly resembling the middle of the minority band gap which is partly occupied by thermally activated states~\cite{Lezaic06}. Both spectra and SP change dramatically at the Fermi energy when p-polarized light is used for excitation. In this case the minority channel exhibits a prominent peak directly at $E_F$, leading to a distinct negative SP of -20\%. This peak can be ascribed to a minority surface state with $\Delta_1$ symmetry for three reasons: (i) it can only be excited by p-polarized light;  (ii) several theoretical works~\cite{Hashemifar05, Wuestenberg12} predict minority surface states for CMS(001) and have been  already confirmed experimentally  for off-stoichiometric CMS in our previous work~\cite{Wuestenberg12}; (iii) the state vanishes when the surface is covered by MgO, as described in detail in the following. Furthermore the majority electron spectra show also a non-vanishing photoemission yield near $E_F$, which we attribute to a mixture of additionally excited $\Delta_1$ bulk states and surface resonances.
\begin{figure}
\centering
\includegraphics[width=\linewidth]{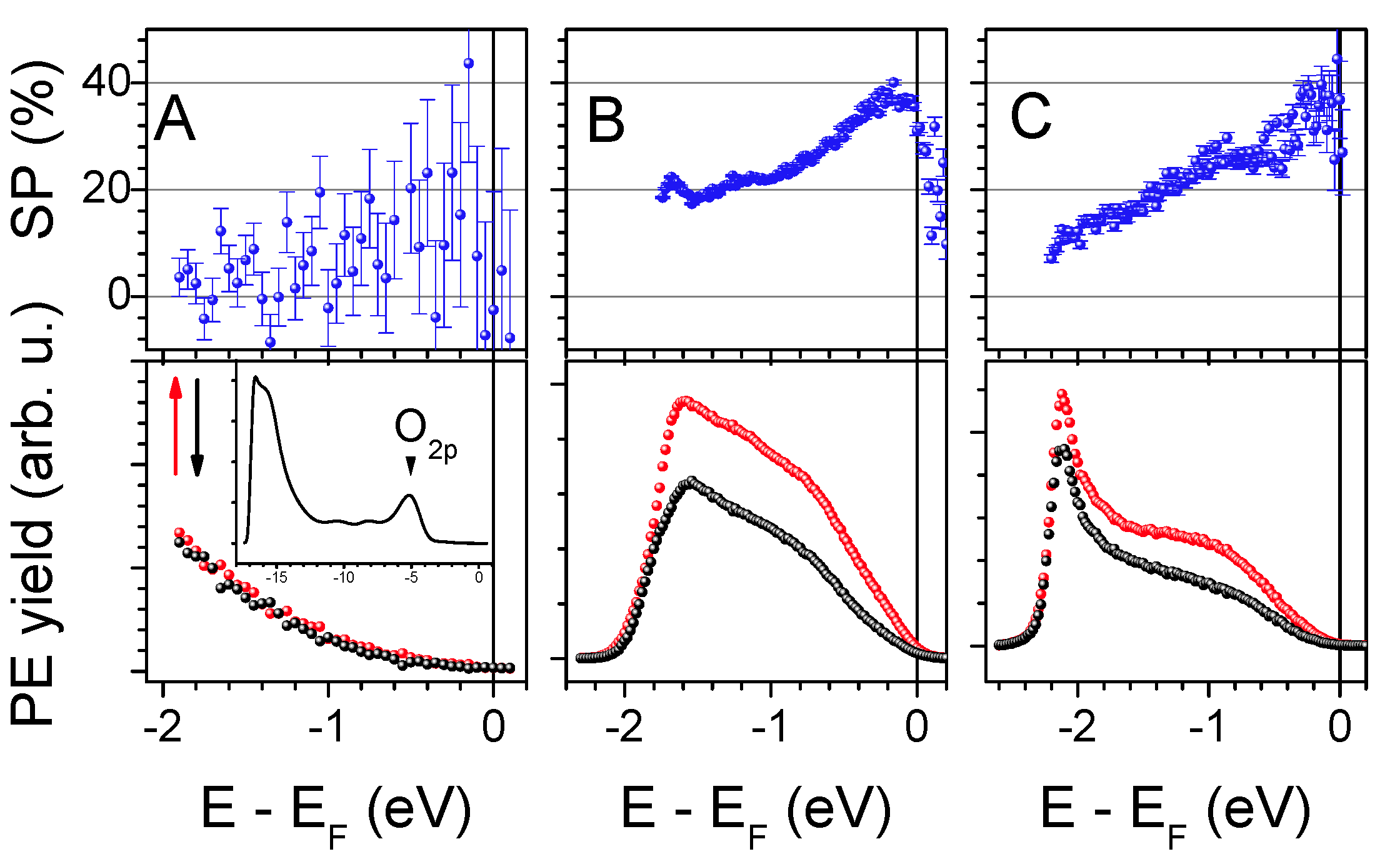}
\caption{Majority (red) and minority (black) electron photoemission spectra of the CMS/MgO(10\,ML) sample (lower panel) and deduced spin polarization (upper panel), measured with SR-UPS (A) and low-energy SR-PES with p-polarized light (B). (C) same as (B), but for the CMS/MgO(20\,ML) sample. Inset in (A): full UPS spectrum of the CMS/MgO(10\,ML) sample with spectral feature from the oxygen 2p peak at  $E_{B}=-5~eV$.}
\label{fig:interfaceSP}
\end{figure} 
\\Let us now turn to the results obtained for the CMS/MgO(10\,ML) system. Figure \ref{fig:interfaceSP}~(A) shows the UPS spectra (bottom panel) and the corresponding spin-polarization (upper panel). In the UPS spectra  almost no photoemission signal is detected close to $E_F$. For lower binding energies a clear feature at $E_{B}=-5~eV$ arising from the oxygen 2p peak in MgO is observed (inset of Fig. \ref{fig:interfaceSP}~(A)). This clearly shows that only the MgO top layer is spectroscopied by SR-UPS, revealing the MgO band gap in the vicinity of $E_F$.    
The low-energy SR-PES results shown in Fig. \ref{fig:interfaceSP}~(B) show a completely different behavior: The spectra themselves resemble the ones obtained at the CMS surface using s-polarized laser light (c.f.  Fig. \ref{fig:surfaceSP}), although the features are  washed out. 
The inferred interface SP also shows high similarity with a maximum value of +~35-40\% in vicinity of the Fermi energy, dropping down to roughly 20\% at higher binding energies. The spectra recorded with s-polarized light (not shown)  are almost identical to the spectra in Fig. \ref{fig:interfaceSP}(B), even though for p-polarization the photoemission yield is increased in general due to excitation of additional states, as will be discussed in detail in the next paragraph. 
The striking similarity of the low-energy SR-PES spectra from the bare CMS surface to those from the CMS/MgO(10\,ML) interface  clearly indicates that the low-energy SR-PES is extremely sensitive to the buried CMS/MgO interface, in strong contrast to SR-UPS. Note that the uncovered CMS sample features bulk-like spectroscopic properties if s-polarized light is used, since surface resonances and surface states can not be excited. Evaporation of MgO on top otherwise leads to a complete suppression of CMS surface contributions, thus no surface-related states can be observed anymore even with p-polarized light.  The small spectroscopic discrepancies between the free CMS surface and the interface originate solely from interface spectra smearing due to both quasi-elastic and inelastic scattering at defects directly at the interface. These defects are inevitably induced by the finite lattice mismatch between CMS and MgO~\cite{Miyajima09, Fetzer12a}. Furthermore our results clearly show that the found $\Delta_1$ minority surface state does not convert into an interface state, as it is the case e.g. at the CoFe/MgO interface~\cite{Bonell12}. This would significantly influence the TMR properties since $\Delta_1$ states by far contribute most to the tunneling current~\cite{Butler01}. We do not find any indication of interface states having $\Delta_1$ or$\Delta_5$ symmetry at least in  the investigated energy range, since no additional features appear in the interface spectra compared to the free surface. \newline
Despite the astonishing fact that we are able to detect an actual interface spin polarization throughout a MgO thickness of 10\,ML, it is further remarkable that the total photoemission yield of the CMS/MgO structure is not lowered distinctively compared to the free CMS surface. 
This observation still holds even for MgO layers thicker than 10\,ML, as demonstrated by the low-energy SR-PES spectra measured from a CMS/MgO(20\,ML) sample shown in 
Fig. \ref{fig:interfaceSP}~(C).\begin{figure}
\centering
\includegraphics[width=\linewidth]{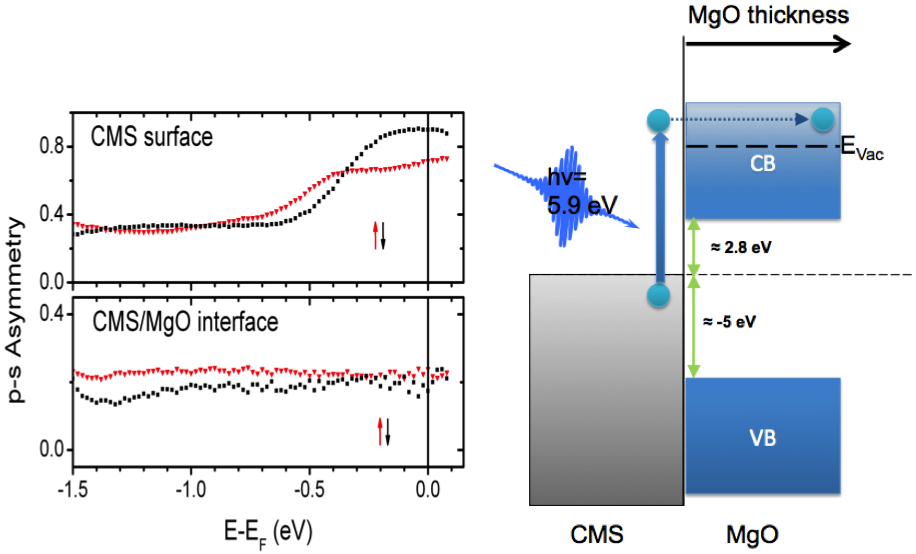}
\caption{Left panel: Energy-resolved p-s asymmetry of the CMS free surface (upper panel) and the CMS/MgO(10\,ML) interface (lower panel). The majority (minority) channel is marked with red triangles (black squares). Right panel: Energy level alignment at the CMS/MgO interface and electrons excited by low-energy SR-PES traversing the MgO layer without scattering, resulting in the interface sensitivity of this method.}
\label{fig:asymmetry}
\end{figure}
The origin of such extreme interface sensitivity can be understood within the following multi-step model:  Low-energy SR-PES, if applied to metals is strongly surface sensitive\cite{Wuestenberg12}, as the electron mean free path is only $\le 1\,nm$. If applied to the CMS/MgO (or more generally to a metal/insulator) interface, due to the high penetration depth of the used laser wavelength in MgO, electrons are excited easily near $E_F$ in the uppermost CMS layers and will traverse the interface. Inside the MgO these electrons will occupy states in the conduction band at 2.8~eV above $E_F$, assuming a band gap of 7.8~eV~\cite{Roessler67} and a valence band maximum at -5~eV~\cite{Fecher08}. Since the excitation energy is 5.9~eV, only the lowest 3.1~eV of the conduction band will be accessible, see the right panel of Fig.~\ref{fig:asymmetry}.   Comparing this to the energy band gap value, one concludes that there is simply no allowed phase space for inelastic electron-electron scattering with MgO valence electrons. Therefore  quasi-elastic transport of the electrons to the sample surface and subsequent emittance is possible even for MgO coverages thicker than 20\,ML,  making low-energy SR-PES extremely interface sensitive.\newline
Before passing,  we evaluate the p-s asymmetry $A=\frac{I_p-I_s}{I_p+I_s}$ using the respective energy-resolved normalized photoemission intensities. This allows us to discuss the relative $\Delta_1$/$\Delta_5$ contributions at the Fermi energy for minority and majority channel of both the CMS surface (upper part of Fig. \ref{fig:asymmetry}, left) and the CMS/MgO(10\,ML) interface (lower part), as explained in Ref.~\cite{Wuestenberg12}. At the free surface both spin channels show enhanced asymmetry values near $E_F$ compared to higher binding energies, resulting from the minority surface state and majority surface resonances (surface contributions intrinsically exhibit $\Delta_1$ symmetry). On contrary, the interface asymmetry varies only marginally, which is in accordance with CMS bulk band structure calculations~\cite{Balke06, Miura07} predicting an identical dispersion of the  $\Delta_1$ and $\Delta_5$ majority bands near $E_F$ along the $\Gamma$-X direction we probe. \newline
Although in this study only measurements at RT are presented, our findings finally allow us to illuminate the actual cause for the drastic performance loss at RT of MTJs consisting of CMS and MgO, which is still controversially discussed. At first we assume that the minority states found by us at the interface are induced only thermally. Since the related p-s-asymmetry is virtually identical to the majority one, a possible origin is that the majority $\Delta_1$ and $\Delta_5$ states are projected partly onto the minority channel due to noncollinear interface magnetic moments. This is in fact predicted theoretically by Miura \etal, who investigated the influence of exchange stiffness lowering at the CMS/MgO interface~\cite{Miura11}. Additionally our findings would explain in an elegant way the experimentally found lack of spectral changes in temperature-resolved spin-integrated photoemission measurements at CMS~\cite{Miyamoto09}, since the majority DOS would simply be mirrored in the minority channel for higher temperatures and hence showing a non Stoner-model-like behavior. \newline 
In summary, we have successfully applied low energy SR-PES to study the spin- and symmetry properties of the CMS/MgO interface buried below MgO films with the same thickness used in state-of-the-art MTJ structures, and thus free from  detrimental MgO pinholes appearing for ultra-thin coverage.  A comparison between the interface spectra and the spectra from the CMS free surface revealed the complete suppression of the minority surface state present at the CMS surface, and no additional interface states at the CMS/MgO interface. The interface spectra thus reflect mainly the bulk CMS DOS with a constant $\Delta_1$/$\Delta_5$ ratio and a distinct positive spin polarization, confirming  the latest theoretical explanation of the TMR ratio temperature dependence in state-of-the-art CMS/MgO MTJs. Moreover, we have shown that the strong interface sensitivity of the low energy SR-PES still holds for MgO barriers as thick as 20\,ML, opening the way for the non-destructive characterization of the spin and symmetry properties of buried metal/insulator interfaces.\newline   
This work was financially supported by the DFG Research Unit 1464 ASPIMATT. The work at Hokkaido University was partly supported by a Grantin-Aid for Scientific Research (A) (Grant No. 23246055) from MEXT, Japan.

\end{document}